\documentclass{webofc}
\usepackage[varg]{txfonts}   
\begin{document}
\title{ALICE Overview}
%
%

\author{\firstname{Alessandro} \lastname{Grelli (the ALICE Collaboration)}\inst{1}\fnsep\thanks{\email{a.grelli@uu.nl}}
}

\institute{Utrecht University, Princetonplein 1, 3584CC, The Netherlands
          }

\abstract{%
  An overview of the recent results obtained by the ALICE Collaboration
from the analysis of the pp, p-Pb and Pb-Pb data samples collected during LHC run I and the first half of run II is presented.
}
\maketitle
\section{Introduction}
\label{intro}
The study of Pb-Pb collisions in the LHC energy regime provides insight on the properties of nuclear matter
at extreme conditions of temperature and energy density, where lattice QCD predicts the matter to be in a Quark Gluon Plasma (QGP) state~\cite{qgp}.
Properties of the QGP medium, such as shear viscosity to entropy ratio, electromagnetic radiation and energy loss provide the focus of past, present  and future proposed measurements.
In addition, in recent years, the investigation of the properties of small systems (i.e. pp and p-A collisions) is acquiring an important role. Since the first experimental evidence of long-range near-side angular correlations in high multiplicity pp~\cite{CMS1} and p-Pb~\cite{alice_r} interactions a large effort has been made to measure the small system properties as a function of the event multiplicity in order to understand the possible onset of collectivity and its implications.  \\
ALICE~\cite{ALICE} is the dedicated heavy ion experiment at the LHC. It is equipped with excellent particle identification (PID) and tracking capabilities down to very low transverse momentum ($p_{\rm T}$). The aforementioned properties make ALICE unique in measuring identified particles spectra, (anti-)(hyper-) nuclei production and in reconstructing decays of heavy-flavour hadrons by resolving their secondary decay vertices both in small systems and in heavy-ion collisions.\\
In the next sections, an overview of recent ALICE results from LHC run I and II data is presented with particular focus on strangeness, hard-probes, nuclei production and collectivity. It has to be noted that, at the moment this manuscript is being written, the LHC run II is still ongoing. At present ALICE collected an integrated Pb-Pb luminosity of $\sim$ $250~\mu b^{-1}$ during the 2015 data taking while additional $\sim$ $750~\mu b^{-1}$ are expected by the end of 2018 for a total run II Pb-Pb integrated luminosity of $L_{\rm int}\sim1~nb^{-1}$. \\
Finally, the ALICE Collaboration is preparing the upgrade of the detector, which will take place during LHC long shutdown 2 in 2019, when, among other interventions, the inner tracking system and the readout chambers of the Time Projection Chamber will be fully replaced~\cite{gines}.

\section{Highlights on strangeness production}

Strangeness enhancement (i.e. higher abundance of strangeness per participating nucleon in A-A collisions with respect to pp collisions at the same energy) was one of the first proposed signatures of the formation of the QGP, suggested by Rafelski and Muller in 1982~\cite{strange1}. The effect was indeed observed for the first time at the SPS~\cite{strange2} and then at RHIC~\cite{strange33}. The investigation of strangeness production at RHIC and LHC, revealed several features, such as the baryon to meson anomaly and the enhancement decreasing with the increase in collision energy, which triggered further theoretical investigations~\cite{strange3}. Quark coalescence has been used to explain the intermediate-$p_{\rm T}$ baryon to meson enhancement~\cite{greco}, while statistical hadronization models succeeded in describing hadron yields using a grand-canonical approach~\cite{peter}.  In the statistical model,
the lower production of strangeness in small systems (pp collisions) is understood as the consequence of the small phase-space volume available, and it is referred to as canonical suppression~\cite{supp} .
\begin{figure}[h]
\centering
\includegraphics[width=6.5cm,clip]{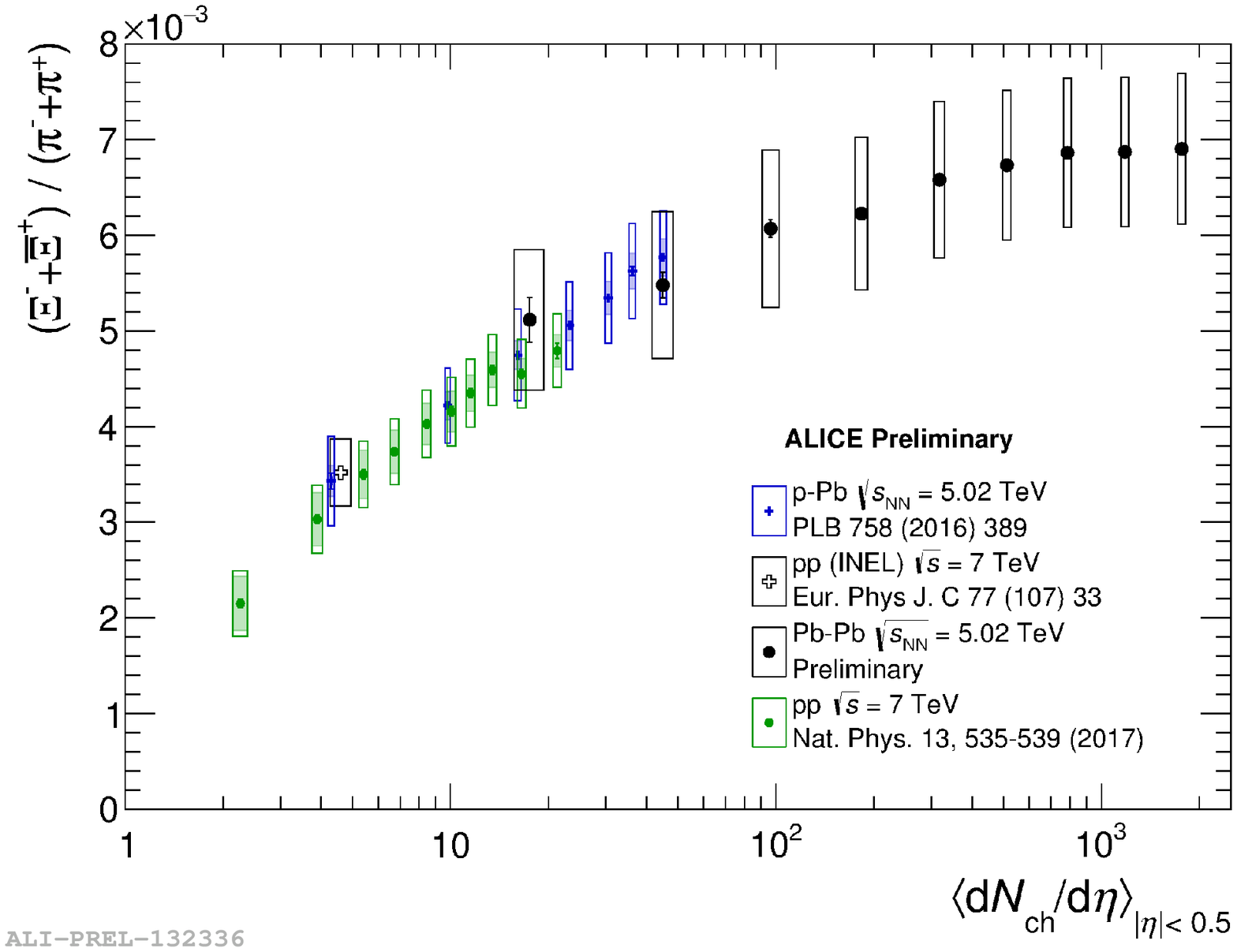}
\includegraphics[width=6.8cm,clip]{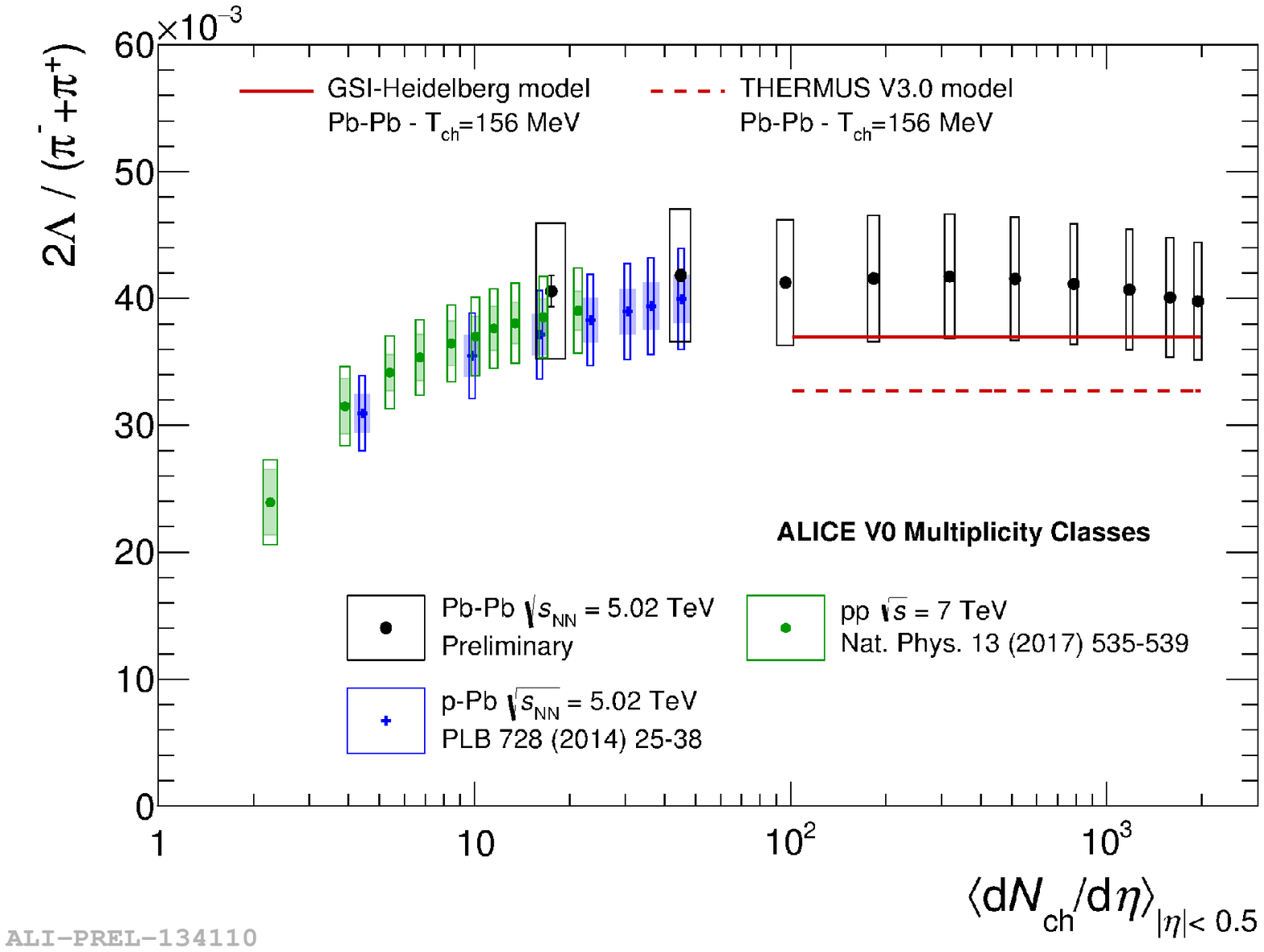}
\caption{Left: Integrated production yield of $\Xi^-+\Xi^+$ normalized to the one of charged pions is show versus the $<\rm dN_{\rm ch}/\rm d\eta>_{|\rm y|<0.5}$. Right: same plot for the $\Lambda$ compared with Thermal Model predictions.}
\label{fig-1}       
\end{figure}
The ALICE experiment detects $K^{0}_{S} , \Lambda, \Xi^{\pm}$ and $ \Omega^{\pm}$ through the reconstruction of the daughter tracks coming from their weak decay in the rapidity region $|\rm y| < 0.5$. The Inner Tracking System and the Time Projection Chamber are used for secondary vertex determination, tracking and PID through specific energy loss measurements.
Recent measurements by the ALICE experiment~\cite{nature} show that strangeness enhancement is also present in high-multiplicity pp and p-Pb events.  Moreover, the $\Lambda / \rm K^0_s$ ratio shows a qualitatively similar behavior in p-Pb and Pb-Pb collisions~\cite{ALICE4}. Therefore, the study of strangeness production in small and large systems acquires a fundamental importance in the investigation of the origin of the heavy-ion-like effects found in small systems~\cite{livio}.\\
The preliminary measurement of the production yield of strange and multi-strange hadrons in Pb-Pb at $\sqrt{s_{\rm NN}}=5.02~{\rm TeV}$ was shown for the first time at this conference. As example, in the left panel of figure~\ref{fig-1}  the integrated production yield of $\Xi^-+\Xi^+$ normalized to the one of charged pions is shown as a function of the charged-particle multiplicity $<\rm dN_{\rm ch}/\rm d\eta>_{|\rm y|<0.5} $ for the three different collision systems. The right panel shows the same ratio for the $\Lambda$ baryon. The results provide clear evidence that in high multiplicity pp interactions strangeness enhancement is comparable with p-Pb and peripheral Pb-Pb.  While the general behavior of the finding can be understood using the grand canonical approach, all the generators and models fail to reproduce the shape and/or the magnitude of the effect~\cite{nature, Michal}. 

\section{Highlights on heavy-flavour production}

\begin{figure}[h]
\centering
\includegraphics[width=7.1cm,clip]{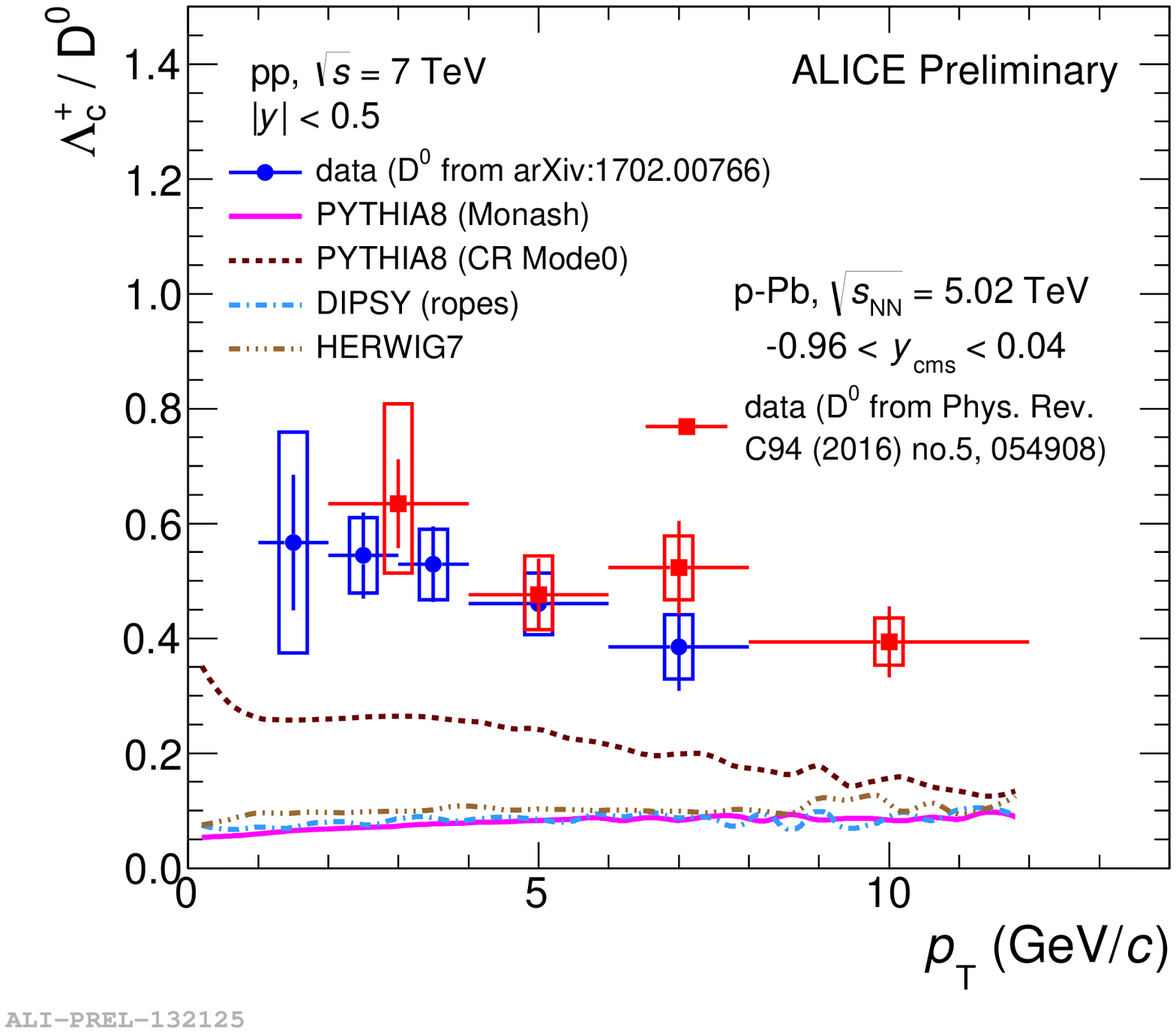}
\includegraphics[width=6.7cm,clip]{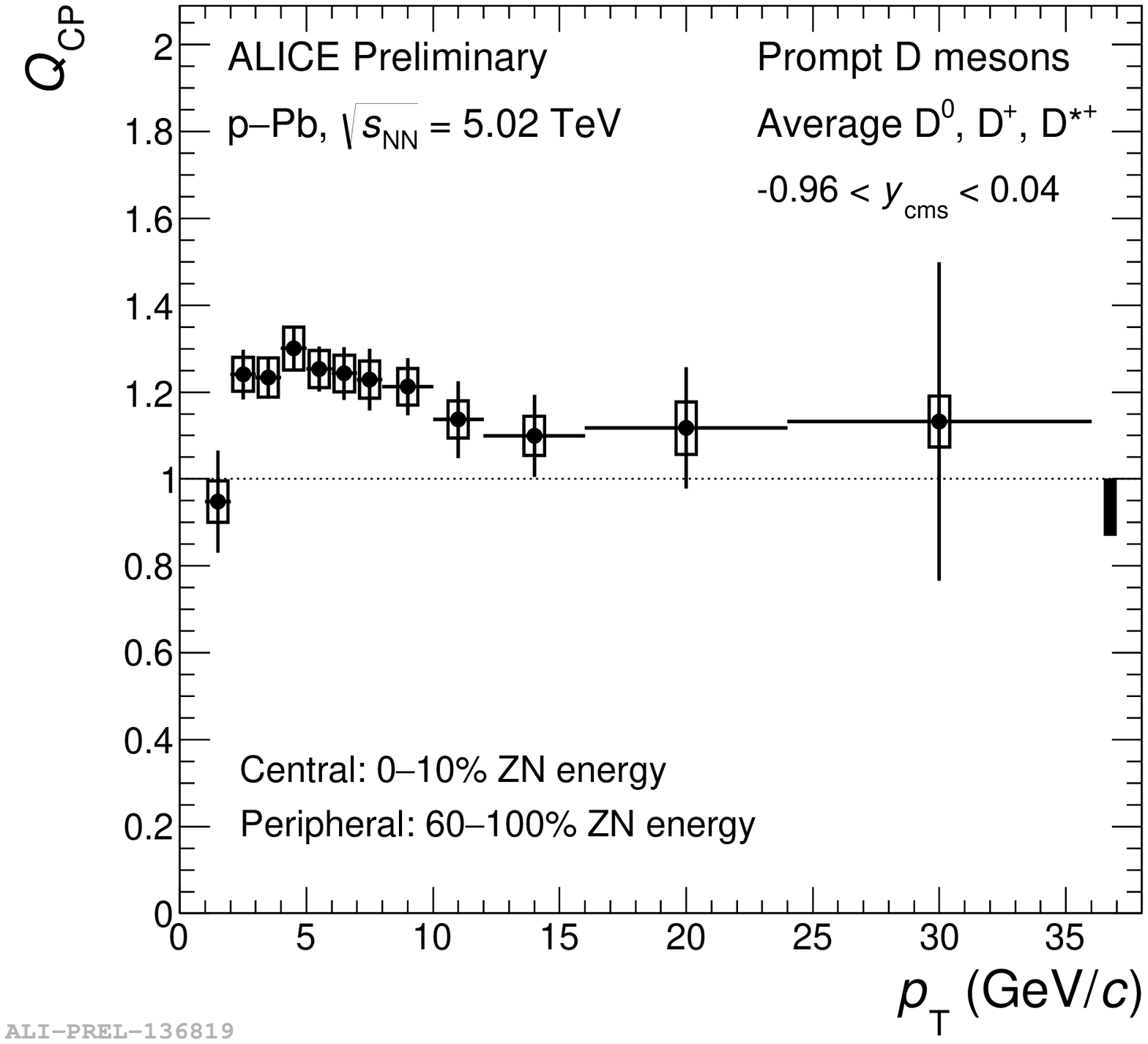}
\caption{Left panel: $\Lambda_c  / \rm D^0$ ratios at central rapidity (|y|<0.5) as measured in pp and p-Pb interactions using LHC run I data at $\sqrt{s} = 7 ~\rm {TeV}$ and $\sqrt{s_{\rm NN}} = 5.02 ~\rm {TeV}$ respectively. Right panel: Average $Q_{\rm CP}$ of non-strange D mesons ($\rm D^0$, $\rm D^+$ and $\rm D^{*+}$) as measured in p-Pb LHC run II data at $\sqrt{s_{\rm NN}} = 5.02 ~\rm {TeV}$. }
\label{fig-2}       
\end{figure}

Heavy-flavour hadrons, which are produced in large-virtuality parton scatterings in the early stages of the collision~\cite{hf1}, are sensitive probes of the medium formed in heavy-ion collisions. They are expected to be sensitive to the energy density through the in-medium energy loss of their heavy quark constituents~\cite{hf2,hf3}. Furthermore, the in-medium thermal production is expected to be negligible~\cite{cinque}. In addition, if in-medium hadronization contributes to charm hadron formation at low-$p_{\rm T}$, then strange charm hadrons (i.e. $\rm D^+_s$ ) should be enhanced relative to non-strange charm hadrons~\cite{sei}. Recently ALICE released the first measurement of two charmed baryon to meson ratios in pp and p-Pb collisions at central rapidity at LHC ($\Lambda_c/\rm D^0$ and $\Xi_c/\rm D^0$)~\cite{Jaime}. Both results challenge models and point toward something not well understood in the fragmentation of charm quark into baryons. The measured $\Lambda_c/\rm D^0$ ratio shown in the left panel of figure~\ref{fig-2} is factor 2-5 higher than all the models considered. On this regard it is interesting to note that only PYTHIA8 with color reconnection can qualitatively reproduce the shape of the measured ratio. The right panel of figure~\ref{fig-2} reports the ratio of the average D meson ($\rm D^0,~\rm D^{*+},~\rm D^+$) production in central to peripheral p-Pb collisions ($Q_{\rm CP}$) obtained at $\sqrt{s_{\rm NN}}$ = 5.02 TeV.  The shape of the ratio shows a hint of deviation from unity at intermediate $p_{\rm T}$ (at $\sim 1.5 \sigma$ level). This deviation could be caused by nuclear modification of the PDFs or by final state effects on D meson production. 

In Pb-Pb collisions the interactions with the produced medium can be studied via the nuclear modification factor, $R_{\rm AA}$, defined as the ratio between the yields measured in Pb-Pb and pp collisions after normalizing the Pb-Pb yield to
the average number of nucleon-nucleon collisions in the considered centrality class. 
\begin{figure}[h]
\centering
\includegraphics[width=5.5cm,clip]{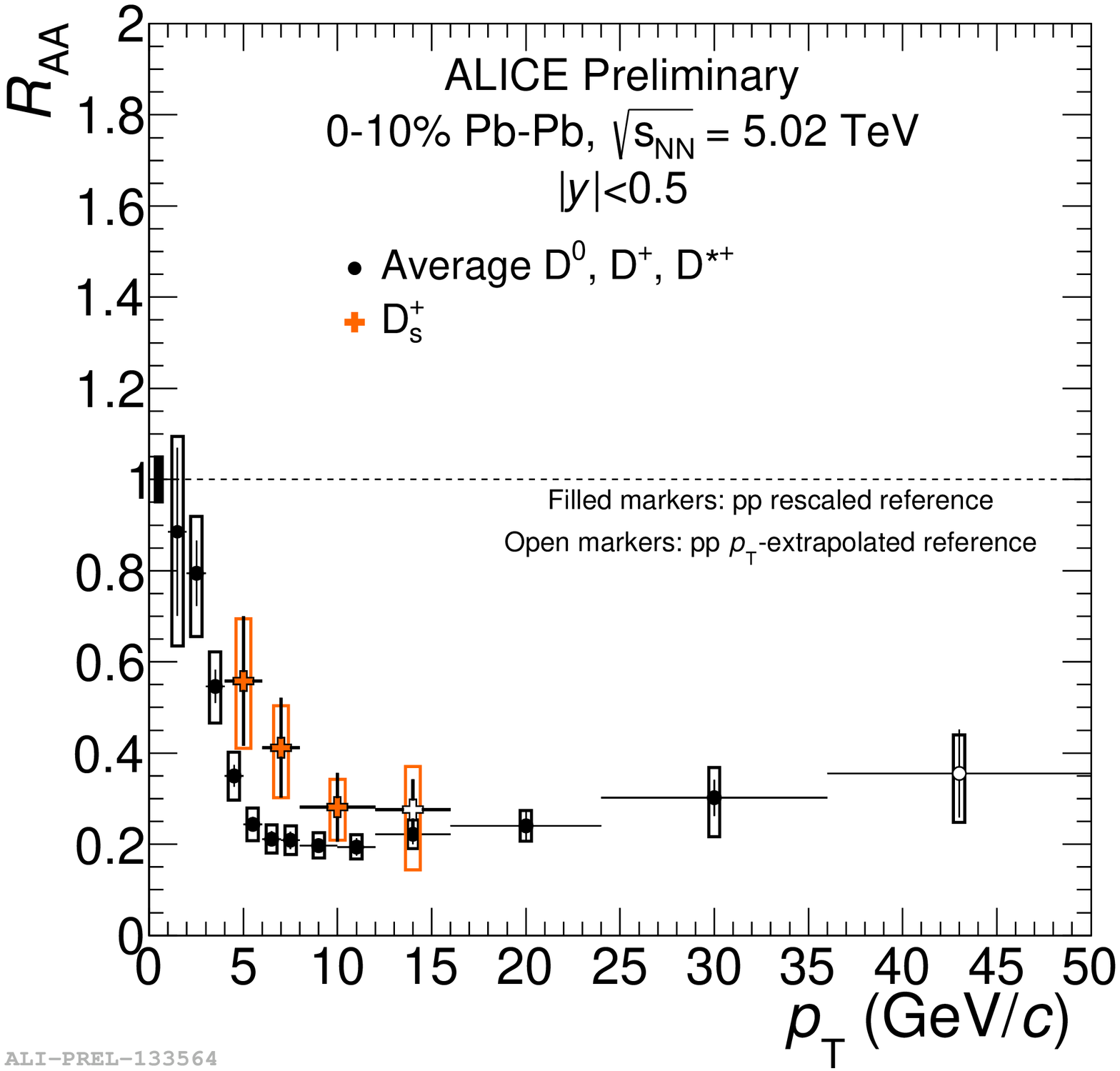}
\includegraphics[width=7.5cm,clip]{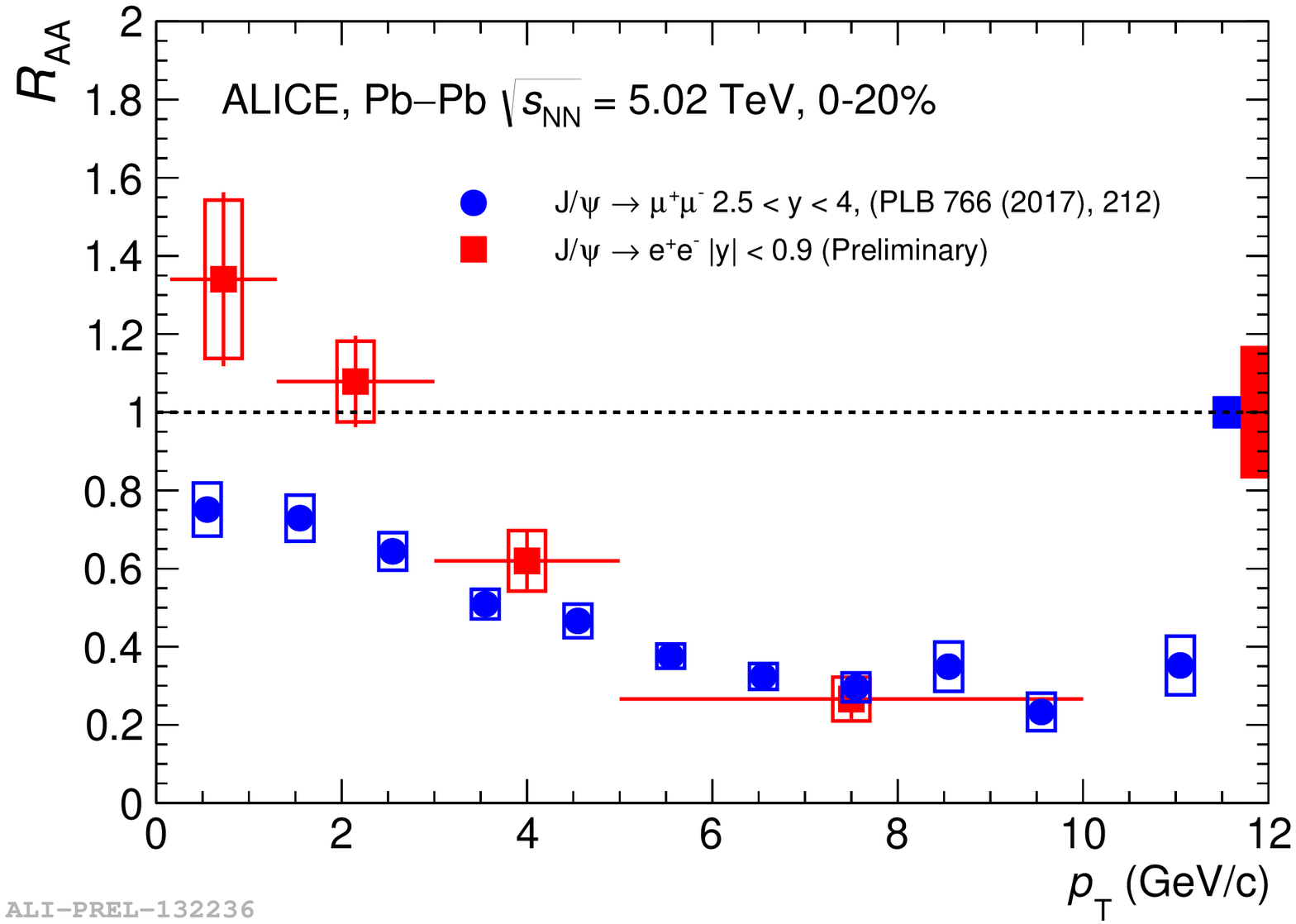}
\caption{Left panel: $R_{\rm AA}$ of non-strange D mesons ($\rm D^0$, $\rm D^+$ and $\rm D^{*+}$) in black compared with the $\rm D^+_s$ $R_{\rm AA}$ in orange as obtained from the most central 10$\%$ Pb-Pb collisions. Right panel: $R_{\rm AA}$ of $\rm J/\Psi$ at central and forward rapidity in the 20$\%$ most central collisions.}
\label{fig-3}       
\end{figure}
It is expected that the medium created in the collision affects the momentum spectra of the originally produced hard probes,
resulting in a value of $R_{\rm AA}$ different from unity. 
It has however to be considered that other effects related to
the presence of nuclei in the initial state can break the expected binary scaling due to the "so called" cold-nuclear matter effects (CNM) (e.g. nuclear modifications of the PDFs, Cronin enhancement, ....). By measuring an $R_{\rm pPb}$ of D mesons compatible with unity in a wide $p_{\rm T}$ range~\cite{Grosa} ALICE suggests that CNM effects are small at the LHC. In the left panel of figure~\ref{fig-3} the non-strange D mesons $R_{\rm AA}$ and the $\rm D^+_s$ $R_{\rm AA}$ are shown for the 10$\%$ most central Pb-Pb collisions. The large suppression found, factor 5-6 at $p_{\rm T}\sim$ 10 GeV/$c$  cannot be explained by CNM effects and therefore is attributed to a final state effect. It has to be noted that the data show hint of relative enhancement of the $\rm D^+_s$ in the low-$p_{\rm T}$ region. However, this observation is not yet significant due to the large uncertainties. \\

Also quarkonium states are expected to be suppressed ($R_{\rm AA} < 1$) in the QGP, due to the color screening of the force which binds the $c\overline{c}$ (or $b\overline{b}$) state. Quarkonium suppression is expected to occur sequentially according to the binding energy of each meson: strongly bound states like   $\rm J/\Psi$ and $\Upsilon ({\rm 1S})$ should melt at higher temperatures with respect to more loosely bound states~\cite{CMSu, ALICEu}.
At LHC energies, it is also predicted that the more abundant production of charm quarks would lead to charmonium
regeneration from recombination of c and $\overline{c}$ quarks at the hadronization level, resulting in an enhancement in the number of observed $\rm J/\Psi$~\cite{costa}.
The right panel of figure~\ref{fig-3}  shows the $R_{\rm AA}$ of $\rm J/\Psi$ measured at central rapidity (|y|<0.9) via the decay channel $\rm J/\Psi \rightarrow e^+e^-$ compared with the same observable measured at forward rapidity (2.5<y<4) via the decay channel $\rm J/\Psi \rightarrow \mu^+ \mu^-$.  The measurement at central rapidity suggests a different behavior at low-$p_{\rm T}$ with respect the forward rapidity measurement and its consistent with the possibility of a stronger regeneration of central rapidity~\cite{Weiser}. However, the large uncertainties of the central-rapidity measurement prevent a clear conclusion on this point.

\section{Highlights on collectivity}

The collective motions arising from the large pressure gradients generated by compressing and heating the nuclear matter is a typical
feature of the medium produced in heavy-ion collisions. The radial flow, generated by the collective expansion of the fireball, is studied by measuring the transverse momentum spectra of identified hadrons \cite{flow1}.
The radial flow velocity at the thermal freeze-out, is estimated via a blast-wave fit. In the Blast Wave model~\cite{bwm}, radial flow breaks the $m_{\rm T}$-scaling present in the non-flowing case. Heavier particles gain more momentum from the common collective flow field and this leads to a mass ordering of elliptic flow at low-$p_{\rm T}$. Indeed a mass ordering in the measurement of elliptic flow of $\pi^{\pm}$, $K^{\pm}, {\rm p} + \overline{\rm p}$, $\phi$, $\Omega^{\pm}$ and $\Xi^{\pm}$ was found by ALICE with data from the first LHC Pb-Pb run at $\sqrt{s_{\rm NN}}$ = 2.76 TeV~\cite{run1flow}. The study also provided evidence for constituent quark scaling~\cite{scaling1,scaling2,scaling3} to be violated at the intermediate $p_{\rm T}$ region at the level of $\pm 20\%$.
\begin{figure}[h]
\centering
\includegraphics[width=14cm,clip]{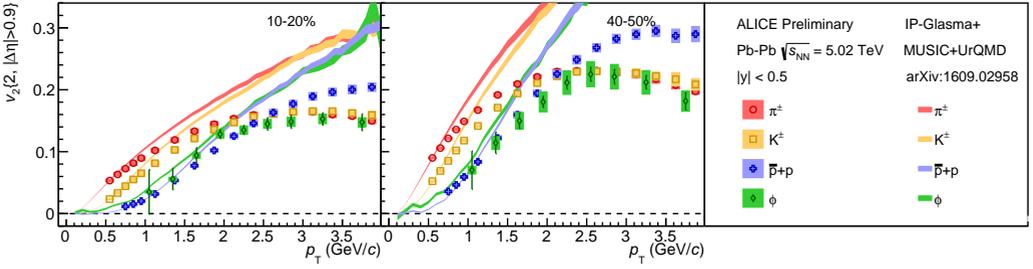}
\caption{$v_{\rm 2}$ for pions, protons, kaons and $\phi$ mesons for various centrality classes, measured with the scalar product method in Pb-Pb at $\sqrt{s_{\rm NN}}$ = 5.02 TeV. Comparison with hydrodynamics simulations is added.}
\label{fig-4}       
\end{figure}
Profiting of the larger sample collected during the second LHC Pb-Pb run at $\sqrt{s_{\rm NN}}$ = 5.02 TeV, ALICE released a new, more precise, measurement of the elliptic flow of $\pi^{\pm}$, $K^{\pm}, {\rm p} + \overline{\rm p}$ and $\phi$ (see Fig. ~\ref{fig-4}) as well as measurement of the higher order harmonics $v_{\rm 3}$ and $v_{\rm 4}$. Hydrodynamical simulations~\cite{hyd} based on ip-Glasma+ MUSIC+UrQMD can reproduce the measured ordering qualitatively rather well in the $p_{\rm T}$ region below 1-2 GeV/$c$ for all the $v_{\rm n}$ considered. 
\begin{figure}[h]
\centering
\includegraphics[width=6.1cm,clip]{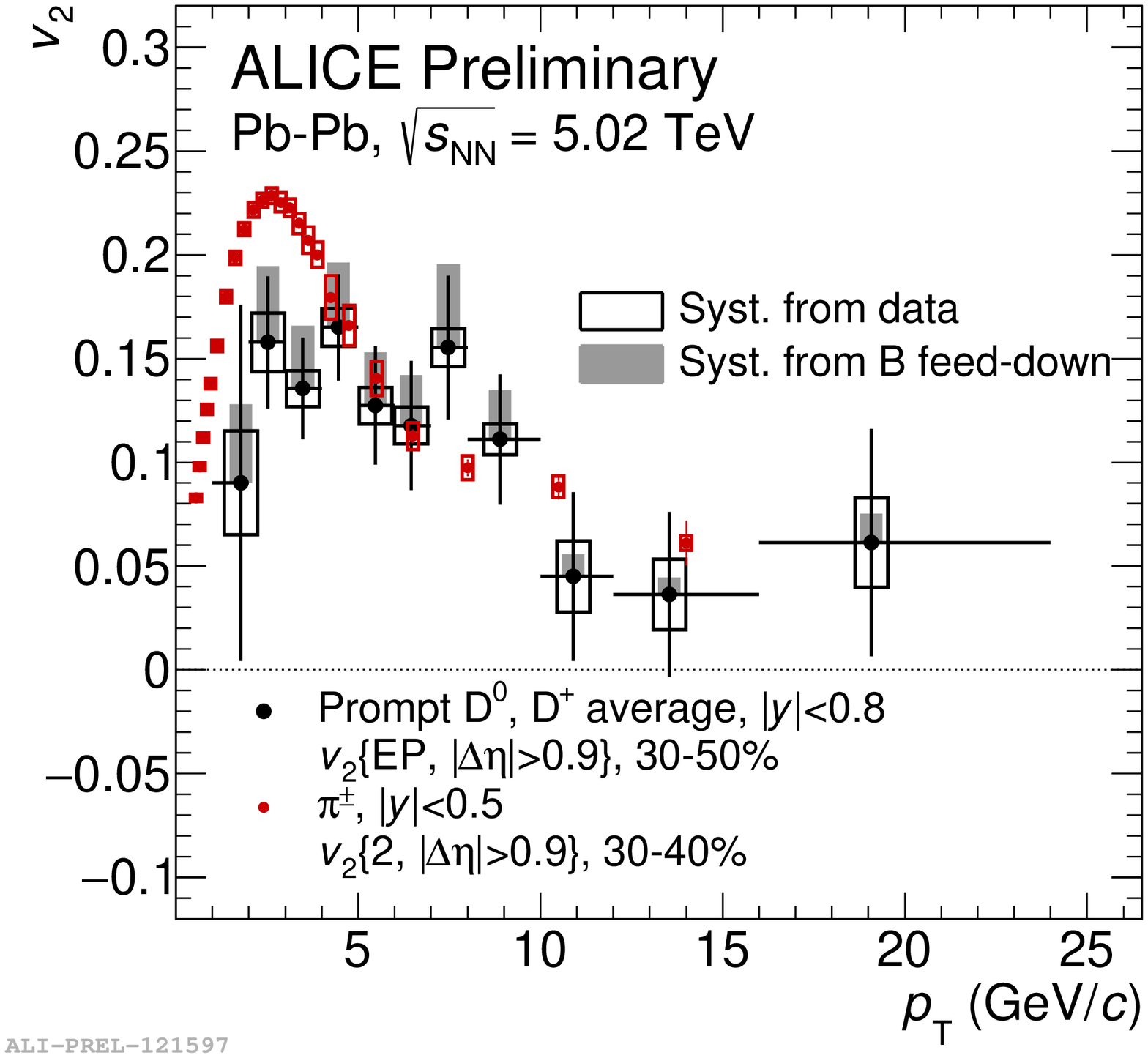}
\includegraphics[width=7.6cm,clip]{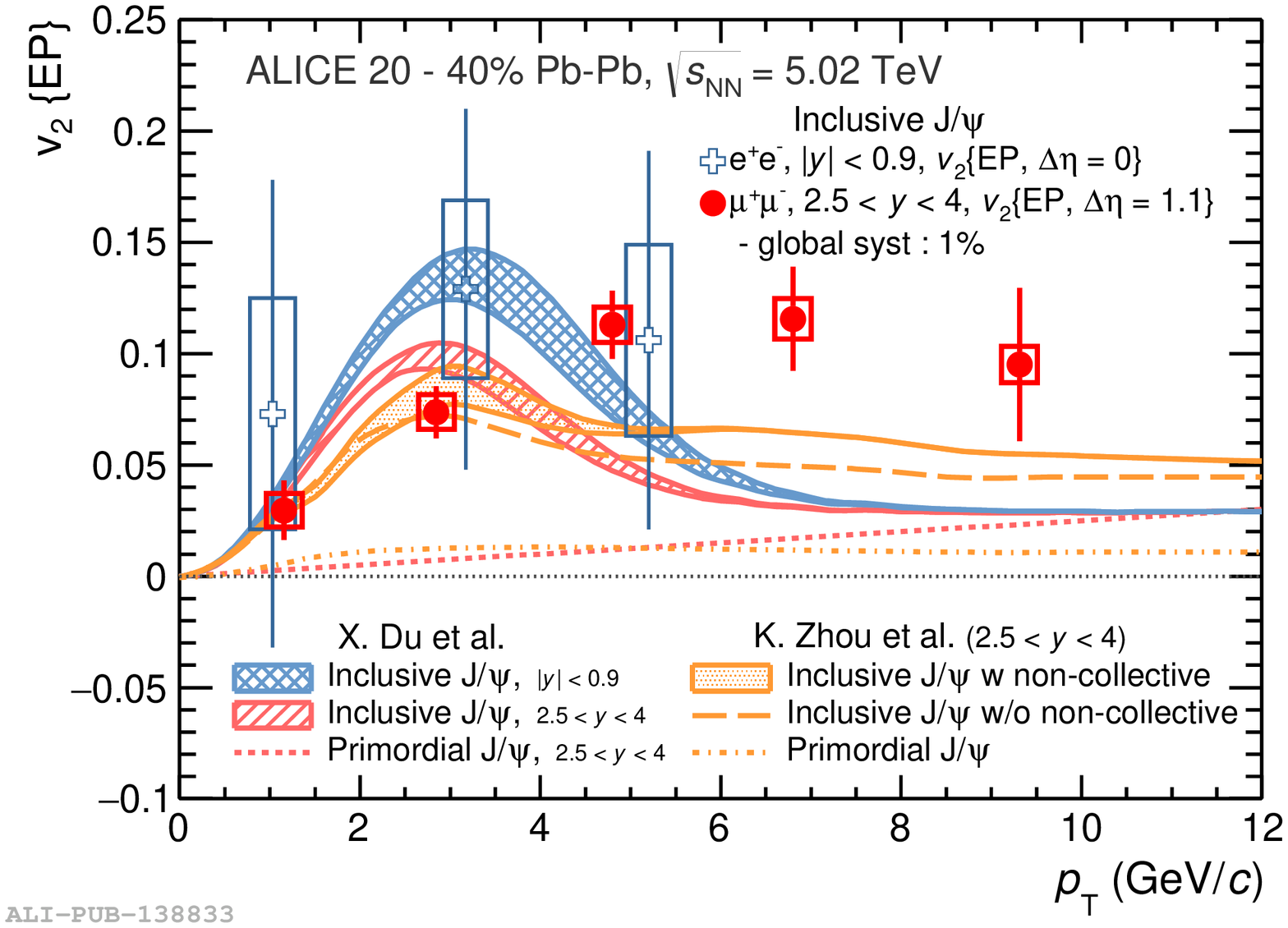}
\caption{Left panel: Average non-strange D-mesons $v_{\rm 2}$ measured in the centrality class $30-50\%$ compared with pion $v_{2}$ measured in the centrality class $30-40\%$. Right panel: ${\rm J}/\Psi$ $v_{\rm 2}$ measured at central ($|\rm y|<0.9$) and forward (2.5<y<4) rapidity.}
\label{fig-5}       
\end{figure}
In figure~\ref{fig-5} on the left panel, the elliptic flow of D-mesons in the centrality range $30-50\%$ , measured with event plane method~\cite{ep}, is compared to charged pion $v_{\rm 2}$. The measurement~\cite{dv2} shows that D-mesons have a positive $v_{\rm 2}$ in range $2 < p_{\rm T} < 10$ GeV/$c$ indicating that charm quark takes part to the collective motion of the system. At higher-$p_{\rm T}$, $v_{\rm 2}$ is compatible with 0 within uncertainties. \\
In the right panel of the same figure the $\rm J/\Psi$ elliptic flow is shown as measured by ALICE at forward ($2.5<\rm y<4$) and central ($|\rm y|<0.9$) rapidities. While the large error bars of the measurement at central rapidity do not allow for a quantitative comparison with theory calculations~\cite{Xu}, for the forward rapidity measurement, at low-$p_{\rm T}$, the theory calculation~\cite{zou} is able to reproduce the magnitude of the measured $v_{ \rm 2}$ by adding a strong $\rm J/\Psi$ regeneration component while at high-$p_{\rm T}$ the magnitude of the $v_{\rm 2}$ is underestimated.

\section{Highlights on (Anti-)(Hyper-) nuclei production}

\begin{figure}[h]
\centering
\includegraphics[width=7.2cm,clip]{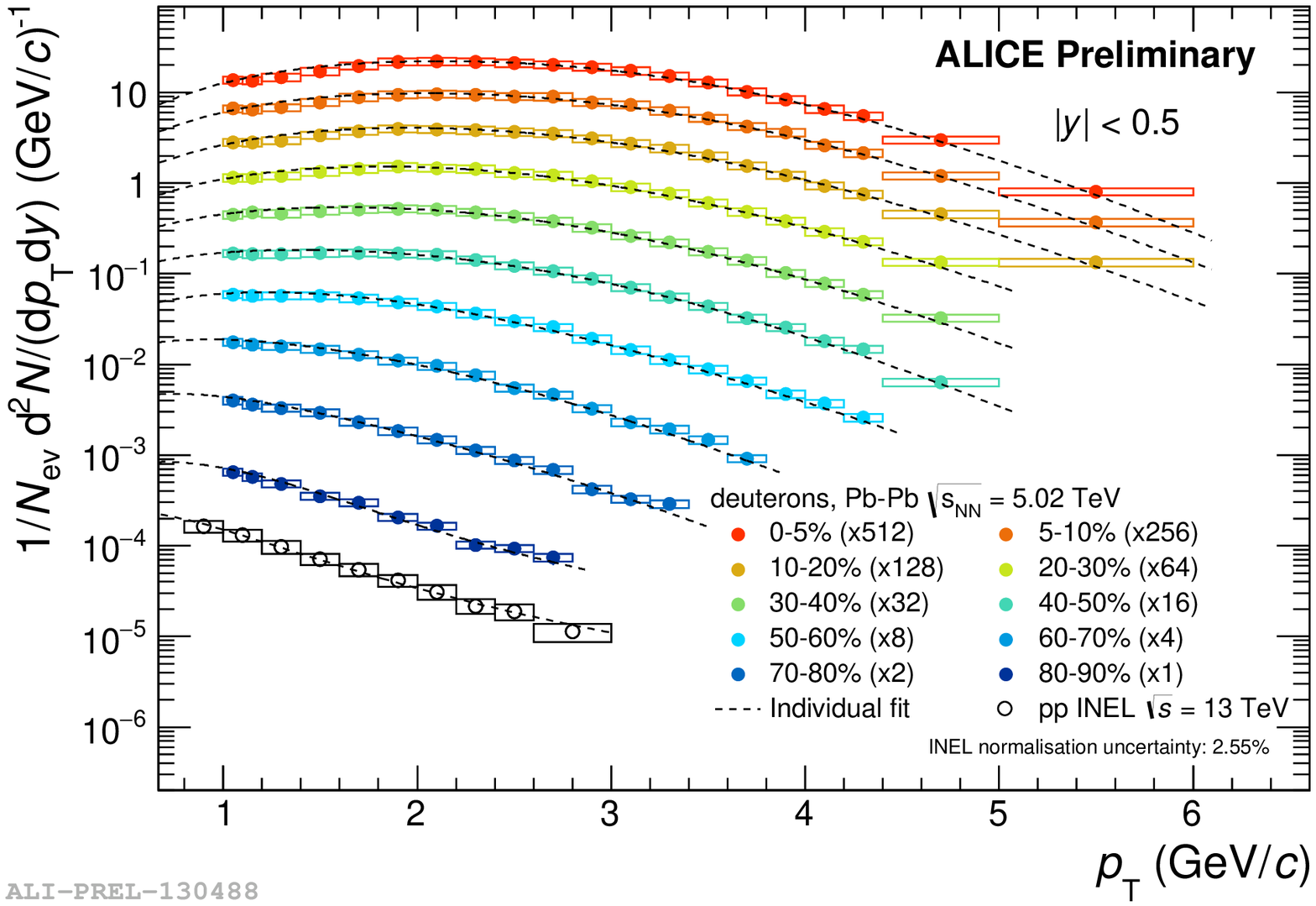}
\includegraphics[width=6.3cm,clip]{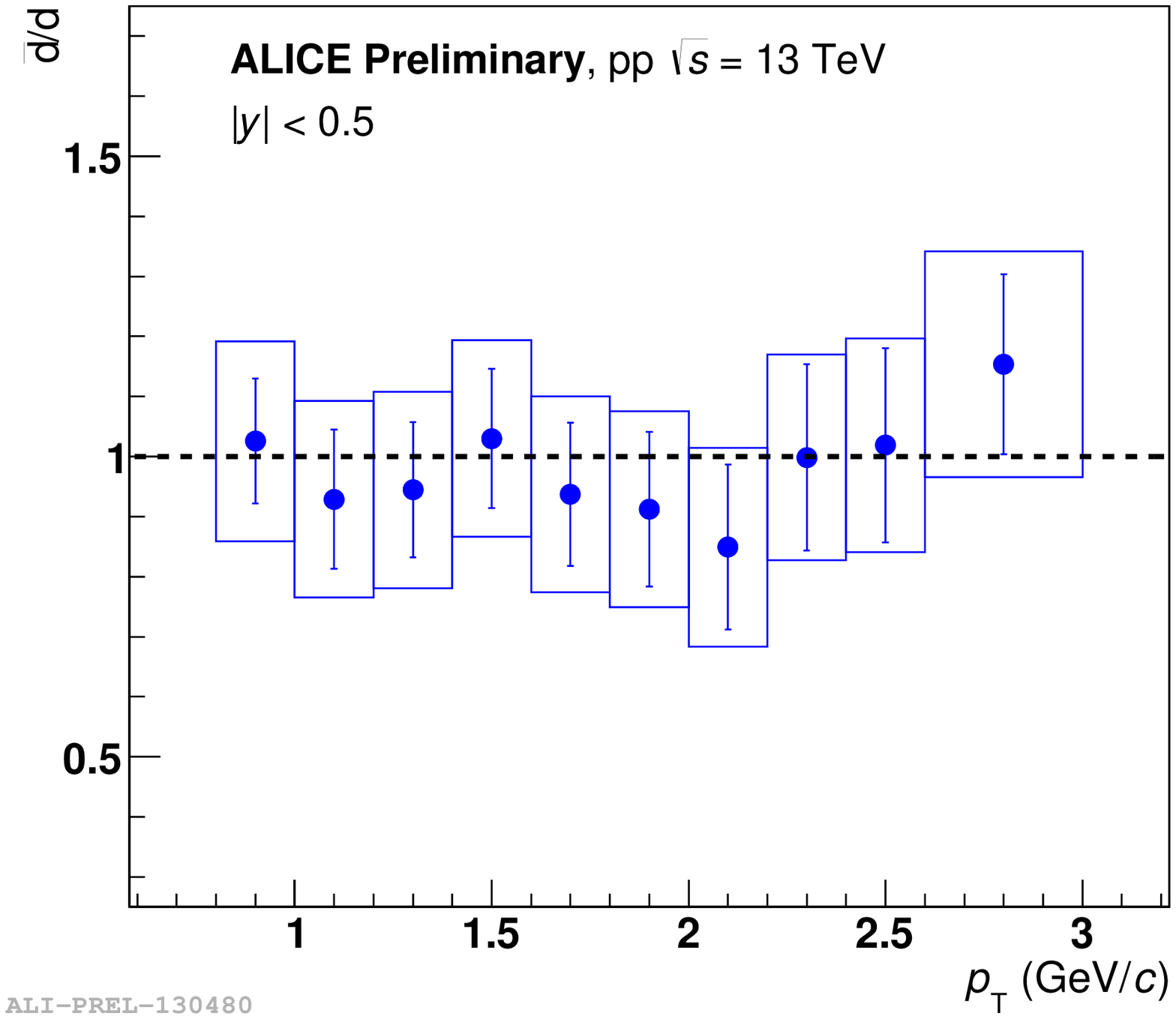}
\caption{Left panel: deuteron production in pp at $\sqrt{s} = 13~{\rm TeV}$ and in Pb-Pb at $\sqrt{s_{NN}} = 5.02~{\rm TeV}$. Right panel: d/$\overline{\rm d}$ ratio in pp at $\sqrt{s} = 13~{\rm TeV}$.}
\label{fig-6}       
\end{figure}

The production of
(anti-)(hyper-)nuclei~\cite{puccio} was measured in all the available collision systems at LHC at central rapidity ($|\rm y|<0.5$).
The production spectra of deuterons in Pb-Pb collisions at $\sqrt{s_{\rm NN}}$ = 5.02 TeV and in pp collisions
at $\sqrt{s}$ = 13 TeV are shown in the left panel of figure~\ref{fig-6}. 
The typical hardening of the spectra with increasing centrality, already observed for lighter particles~\cite{Adam} and for deuterons~\cite{adam2} at lower energies, is visible. The observation suggests an increasing radial flow with increasing centrality. Moreover, in the right panel of the same figure, the ratio between the production spectra of deuteron and and-deuteron is shown to be compatible with unity. This observation is compatible with expectations from the statistical hadronisation~\cite{dieci} and the hadron coalescence models~\cite{undici}.\\ Finally, ALICE presented one of the most precise measurements of the ${^{3}_{\Lambda}\rm H}$ lifetime using the Pb-Pb data sample at $\sqrt{s_{\rm NN}}$ = 5.02 TeV collected in 2015. The preliminary measured value of $\tau = 237^{+33}_{-36} ({\rm stat.}) \pm 17 ({\rm syst.})~ {\rm ps}$ is compatible with both the previously computed world average of $\tau = 216^{+18}_{-16}~ {\rm ps}$~\cite{vita} and with the free $\Lambda$ lifetime.

\section{Conclusions}

ALICE is the dedicated heavy ion experiment at the CERN LHC and it is equipped with excellent tracking and particle identification capabilities. The large sample collected during the second Pb-Pb run at LHC allowed ALICE to enter in a precision hera for many observables (e.g. strangeness production and collectivity in Pb-Pb). In terms of azimuthal anisotropy, measurements of the $v_{\rm 3}$ and $v_{\rm 4}$ of identified particles set additional tight constraints to models. On the heavy-flavour sector ALICE released the first measurement of the charmed baryon to meson ratio in pp and p-Pb interactions at central rapidity at LHC. Such a measurement constitutes, at present, a challenge for models of charm fragmentation.  
There are interesting hints for collective behavior in high-multiplicity pp and p-Pb collision, one of them being the measurement of the $Q_{\rm CP}$ of non-strange D mesons. However, the low significance and the fact that the effect of a Color Glass Condensate or other cold matter effects may yet provide an alternative explanation for the observed behavior make the measurement not yet conclusive.
The excellent PID capabilities enable ALICE to perform high precision nuclei measurements. At this conference ALICE presented precision measurements of deuteron and anti-deuteuteron production in pp and Pb-Pb as well as one of the most precise measurements so far of the hyper-triton lifetime.

%
%
%

\end{document}